\documentclass[aps,pra,twocolumn,letterpaper,superscriptaddress,10pt]{revtex4-2}
\usepackage{amssymb,amsthm,amsmath,amsfonts}
\usepackage{graphicx,ulem,enumerate,bbm,bm}
\usepackage[pdftex,dvipsnames,usenames]{xcolor}
\usepackage[colorlinks=true,urlcolor=blue,citecolor=blue,linkcolor=blue]{hyperref}

\begin{document}

\title{Observable nonclassicality witnesses for multiplexed detection systems}

\author{S. Krishnaswamy}
    \email{suchitra@mail.uni-paderborn.de}
    \affiliation{Paderborn University, Institute for Photonic Quantum Systems (PhoQS), Theoretical Quantum Science, Warburger Stra\ss{}e 100, 33098 Paderborn, Germany}

\author{M. Jung}
    \affiliation{Institute of Condensed Matter Theory and Optics, Friedrich-Schiller-Universit\"at Jena, Max-Wien-Platz 1, 07743 Jena, Germany}

\author{L. Ares}
    \affiliation{Paderborn University, Institute for Photonic Quantum Systems (PhoQS), Theoretical Quantum Science, Warburger Stra\ss{}e 100, 33098 Paderborn, Germany}

\author{M. G\"arttner}
    \affiliation{Institute of Condensed Matter Theory and Optics, Friedrich-Schiller-Universit\"at Jena, Max-Wien-Platz 1, 07743 Jena, Germany}

\author{J. Sperling}
    \affiliation{Paderborn University, Institute for Photonic Quantum Systems (PhoQS), Theoretical Quantum Science, Warburger Stra\ss{}e 100, 33098 Paderborn, Germany}

\date{\today}

\begin{abstract}
	We address the problem of constructing witnesses for nonclassical light that are applicable in state-of-the-art photon-counting devices.
	The key ingredient for the criteria we derive are generalized and directly measurable counting statistics and matrices of counting moments.
	Beyond common criteria, we find classes of witnesses that are based on half-integer powers of click moments and counts.
	Remarkably, this leads to an exponential increase of the number of nonclassicality criteria one can construct and apply.
	With this finding, special attention is payed to probing even and odd parity states, requiring such distinct witnesses.
	Our method is applicable to spatial and time-bin multiplexing in optical systems, where each spatial and temporal mode can be measured with both on-off detectors and detectors with partial internal quasi-photon-number resolution.
	Generalizations to multimode scenarios are provided, allowing for the direct measurement of nonclassical correlations and coincidence counts between an arbitrary number of modes.
\end{abstract}

\maketitle

\section{Introduction}
\label{sec:Introduction}

	Ever since the quantum nature of light was proven \cite{E1905}, there has been an enormous growth of research on this topic, leading to today's technological advancements that harness the particle properties of light \cite{BFV09}.
	Thus, there is a broad interest in quantum-optical states of light \cite{G63}, pertaining to their generation \cite{WKHW86,TRS87,VBW04}, detection, \cite{GM87,LP95}, reconstruction \cite{WVT99,TZERH11,LR09,R96}, and application in quantum metrology \cite{LADDESMPT24} and quantum computing \cite{KLM01,VBR08}.

	Because of the importance of quantum optics for fundamental and applied research, there is a need to classify quantum states of light, separating them into families of classical and nonclassical states \cite{M84}.
	On an abstract level, nonclassicality is defined as the failure of the Glauber--Sudarshan phase-space representation to be interpreted as a non-negative distribution \cite{S63,G63}.
	However, due to well documented problems, such as highly singular quasiprobabilities and the demand for quantum-state tomography (see, e.g., Ref. \cite{SV20} for an overview), a variety of experiment-friendly methods to certify the nonclassicality of light have been devised.
	Such techniques enable one to probe various quantum properties of nonclassical states \cite{OYOWGN25,YXCX19,LZ19,BHPK19,L91,KSKS24}, even extending to multimode coincidence measurements to verify nonclassical correlations \cite{RW84,ZM90}.

	The statistical moments of the bosonic photon-number distribution are often favored for their experimental accessibility \cite{L64,LW66}.
	For instance, the pioneering Mandel parameter relates the variance and mean of a measured photon-number statistics, thereby coining the concept of sub-Poissonian light \cite{M79,M83}.
	However, some states' nonclassicality is not detectable via this parameter \cite{HM85}, requiring extensions of this approach.
	For example, higher-order moments---commonly collected as entries of so-called matrices of moments---go beyond second-order variances \cite{K98,PM06, PMH17,PHM19,AT92,SRW05,APHM16} and extend to multimode scenarios \cite{SW05a,MPHH09,MBWLN10}.
	Another example is Klyshko's criterion \cite{K96}, which can be formulated in terms of the measured photon statistics, circumventing the need to compute moments.
	To account for the limitations in experiments, such as losses and resolution limits, re-designed versions of the Klyshko criteria have been successfully derived \cite{QMCGCVBGLF18,ILF22}.

	The aforementioned criteria require the photoelectric detection of light \cite{KK64}.
	Thus, photo-detectors are essential for determining nonclassicality, emphasizing the need for well-characterized measurement devices \cite{LS99,LFCPSREPW09,B98,DMR77,BKSSV17}.
	Beginning with photo-multiplier tubes \cite{P72}, many kinds of detectors have been employed, such as avalanche photo-diodes (APDs) \cite{J64,KKMBYM81}.
	Today, the most commonly used photo-detection devices are superconducting nanowire single-photon detectors (SNSPDs) \cite{SGK01,VKWLBCZWBAVZSKBMNS21,SSSSSBSB25} and transition-edge sensors (TESs) \cite{I95,IH05,LMN08,HSGSSRBHR19}.

	Key factors determining the performance of a photo-detector are its detection efficiency, dark-count rate, and resolution \cite{H09}.
	Most detectors can only distinguish between the presence of incident photons and their absence, dubbed click and no-click counts, irrespective of the number of incident photons \cite{CGLRZ04,LYLT25,M04}.
	Therefore, a Poissonian photoelectric counting distribution gives an incorrect statistical picture, for example, resulting in false positives when witnessing nonclassicality via the Mandel parameter \cite{SVA12}.
	This dilemma was partially resolved by methods, such as optical multiplexing \cite{PTKJ96,ASSBW03,KW03,RHHPH03}, allowing to improve resolution, while loss deconvolution techniques help with retrieving unattenuated stastical information \cite{STG08}.
	Click-counting theory provides an even more accurate statistical picture \cite{SVA12b,KSADSBSS24} as it takes the binomial nature of the click counts into account.

	Utilizing the experimentally accessible click counts, click-counting moments and the related nonclassicality criteria via matrices of moments can be constructed \cite{SVA13}.
	For example, the notion of sub-binomial light has been introduced and applied to this end \cite{SVA12,BDJDBW13,HSPGHNVS16}.
	Further developments in photo-detection technologies also enabled improving the photon-number resolution of detectors, allowing them to count more than the presence and absence of photons \cite{SLHSSBSB24,SKGCS23, ZHZDYXLLWYXL24}, which can be described as a detector-intrinsic pseudo-photon-number resolution.
	Combined with other methods, such as multiplexing, one can achieve a high pseudo-photon-number resolution \cite{EHBAGDCP23,HGDJ24}.
	This also yields alternative ways to interpret nonclassicality, for example, through the notion of sub-multinomial light \cite{SECMRKNLGWAV17,SCEMRKNLGVAW17,SPBTEWLNLGVASW20}.

	In this paper, we derive previously unknown nonclassicality criteria, using the unconventional approach of harnessing half-integer-power moments as well as half integer counts of multiplexed click-counting devices.
	We apply this approach to multiplexed detectors with and without intrinsic detector resolution.
	We compare our strategy with common approaches, limited to integer-power moments and full counts. 
    We demonstrate the versatility and added benefit of the half-integer technique which also reveals interesting physical relations to the photon-number parity of quantum states.

	The paper is structured as follows:
	In Sec. \ref{Sec:PhotonCounts}, we introduce our half-integer paradigm for the case of photon-number detectors, formulating generalized Klyshko-like and higher-order matrices of moments witnesses for nonclassicality.
	Generalizations to optical multiplexing of binary on-off detectors and multinomial detectors with intrinsic resolution are derived in Secs. \ref{Sec:ClickCounts} and \ref{Sec:MultiClickCounts}, respectively.
	Extensions to measurements for an arbitrary number of modes can be found in Sec. \ref{Sec:Multimode}.
	Eventually, Sec. \ref{Sec:Conclusion} summarizes this work and provides an outlook.

\section{Preliminaries}
\label{Sec:PhotonCounts}

	Before we begin with multiplexing layouts that are used to provide pseudo-photon-number resolution, we revisit the case of ideal photoelectric counting theory.
	This serves as a benchmark and inspiration for the advanced framework formulated later in this work.
	Specifically, the role of parity-based assessments of nonclassicality is highlighted here, yielding criteria that are constructed via half-integer photon numbers and moments.

\subsection{Basic criteria for photoelectric counting}

	In an ideal, single-mode measurement, the photoelectric counting theory is based on a quantum version of the Poisson statistics \cite{MW66}.
	Namely, the photon-number distribution reads
	\begin{equation}
		p_n
		=
		\left\langle{:}
			\frac{\hat n^n}{n!}e^{-\hat n}
		{:}\right\rangle,
	\end{equation}
	where $\hat n$ is the photon-number operator, $n\in\mathbb N$ is the number of detected photons, and ${:}\cdots{:}$ denotes the normal-ordering prescription \cite{AW70}.
	In addition, the normally ordered moments of the photon-number distribution can be determined as factorial moments \cite{G63a},
	\begin{equation}
		\left\langle{:}
			\hat n^m
		{:}\right\rangle
		=
		\sum_{n=m}^\infty n(n-1)\cdots (n-m+1) p_n.
	\end{equation}

	Classical light of a quantized radiation field is described as a statistical mixture of coherent states.
	It is well known that, for any operator $\hat f$, classical states obey the relation \cite{VW06,A12}
	\begin{equation}
		\label{eq:GeneralWitness}
		0
		\stackrel{\text{cl.}}{\leq}
		\left\langle{:}
			\hat f^\dag\hat f
		{:}\right\rangle
		.
	\end{equation}
	A violation of this relation is a witness of a nonclassical state of quantum light.

	Choosing a suitable $\hat f$ is the paramount for harnessing all the information encoded in experimentally obtained data \cite{V00,RV03,MBWLN10}.
	A common example is related to the matrix of moments.
	As a first instance, one can make the choice
	\begin{equation}
		\hat f=\sum_{k\in \mathcal{I}} f_k \hat n^k,
	\end{equation}
	for $f_k\in\mathbb C$ and $\mathcal{I}$ being an index set that is going to be carefully determined later.
	The classicality constraint in Eq. \eqref{eq:GeneralWitness} takes the form
	\begin{equation}
		\label{eq:WitnessMoments}
		0
		\stackrel{\text{cl.}}{\leq}
		\left\langle{:}
			\hat f^\dag\hat f
		{:}\right\rangle
		=
		\sum_{k,l\in \mathcal{I}} f^\ast_k f_l
		\left\langle{:}
			\hat n^{k+l}
		{:}\right\rangle
		=\vec f^\dag M \vec f,
	\end{equation}
	with a vector $\vec f=[f_k]_{k\in \mathcal{I}}$ and the so-called matrix of moments \cite{AT92,SRW05}
	\begin{equation}
     \label{eq:MomentMatrix}
		M
		=
		\left[
			\left\langle{:}
				\hat n^{k+l}
			{:}\right\rangle
		\right]_{k,l\in \mathcal{I}}.
	\end{equation}
	The statement in Eq. \eqref{eq:WitnessMoments} for arbitrary $\vec f$ is equivalent to a positive semidefinite matrix of moments,
	\begin{equation}
		\label{eq:PosSemiDefMoments}
		0
		\stackrel{\text{cl.}}{\leq} M.
	\end{equation}
	Again, $M\ngeq0$ certifies nonclassical light.

	Slightly more involved, but functioning in the same manner, are criteria based on the photon-number distribution $p_n$ itself.
	We can take
	\begin{equation}
		\hat f
		=
		\sum_{k\in \mathcal{I}}
		f_k
		\hat n^ke^{-\hat n/2}.
	\end{equation}
	By substituting this choice into Eq. \eqref{eq:GeneralWitness}, we find
	\begin{equation}
		\label{eq:WitnessCounts}
		0
		\stackrel{\text{cl.}}{\leq}
		\left\langle{:}
			\hat f^\dag\hat f
		{:}\right\rangle
		=
		\sum_{k,l\in \mathcal{I}} f^\ast_k f_l
		(k+l)!\,p_{k+l}
		=\vec f^\dag C \vec f,
	\end{equation}
	allowing us to define a matrix of counts
	\begin{equation}
    \label{eq:ClickCountsMatrix}
		C
		=
		\left[
			(k+l)!p_{k+l}
		\right]_{k,l\in \mathcal{I}}.
	\end{equation}
	Analogous to the above discussion, we can state that Eq. \eqref{eq:WitnessCounts} is equivalent to $C$ being positive semidefinite for classical light,
	\begin{equation}
		\label{eq:PosSemiDefCounts}
		0
		\stackrel{\text{cl.}}{\leq}
		C,
	\end{equation}
	and $C\ngeq 0$ certifies nonclassical light.
	See also Refs. \cite{K96,K96a} in this context.

\subsection{The index set and photon-number parity}

	What is often overlooked is the role of the index set $\mathcal{I}$ in the aforementioned nonclassicality witnesses, Eqs. \eqref{eq:WitnessMoments} and \eqref{eq:WitnessCounts}.
	For both the matrix $C$ of photocounts and the matrix $M$ of factorial moments, we ought to have $k+l\in\mathbb N=\{0,1,2,\ldots\}$ to have well defined entries $p_{k+l}$ and $\langle{:}\hat n^{k+l}{:}\rangle$, respectively.
	Importantly, this means that $k+l\in\mathbb N$ must hold true for all $k,l\in \mathcal{I}$.

	Let us explore the previous statement in more detail.
	Firstly, for $k=l$, we have $k+l=2k\in\mathbb N$.
	This implies that $k\in \frac{1}{2}\mathbb N=\{0/2,1/2,2/2,3/2,\ldots\}$ holds true, meaning that all elements of $\mathcal{I}$ are either integers or half-integers.
	Secondly, for $k\neq l$, the condition $k+l\in\mathbb N$ means that either both $k$ and $l$ are integers or both are half integers to ensure in either case that the sum $k+l$ is an integer.
	Thus, we have the following two mutually exclusive options for the choice of the index set:
	\begin{equation}
		\label{eq:SelectionIndexSet}
		\text{either}\quad
		\mathcal{I}\subseteq\mathbb N
		\quad\text{or}\quad
		\mathcal{I}\subseteq\frac{1}{2}\mathbb N\setminus\mathbb N.
	\end{equation}
	In other words, the above criteria must be formulated for half-integers and integers separately;
    and those are all possible options that obey the above constraints on the index set.
	In particular, the half-integer option is often not accounted for and is going to result in larger classes of nonclassicality witnesses when applied to multiplexed click-counting techniques.
	We further remark that expressions such as $\hat n^{1/2}$ denote the positive branch of the square root throughout this paper, e.g., $\hat n^{1/2}=\sum_{n=0}^\infty \sqrt{n}|n\rangle\langle n|$ in spectral decomposition, where $\sqrt{n}\geq0$ is used, which is important for the correct performance of our approach as a nonclassicality witness.

	As an example, say the set $\mathcal{I}$ has only two elements.
	The non-negativity of the matrices $C$ and $M$ for classical light can be related to non-negative determinants by Silvester's criterion via
	\begin{subequations}
	\begin{align}
		\label{eq:CountingBasedExample}
		0
		\stackrel{\text{cl.}}{\leq}{}&
		\det C
		=
		(2k)!(2l)!\,p_{2k} p_{2l}
		-(k+l)!^2\,p_{k+l}^2,
		\\
		\label{eq:MomentBasedExample}
		0
		\stackrel{\text{cl.}}{\leq}{}&
		\det M
		=
		\left\langle{:}
			\hat n^{2k}
		{:}\right\rangle
		\left\langle{:}
			\hat n^{2l}
		{:}\right\rangle
		-
				\left\langle{:}
			\hat n^{k+l}
		{:}\right\rangle^2.
	\end{align}
	\end{subequations}

	If $k,l\in\mathbb N$ holds true, then the above criteria specifically address nonclassicality of states with an odd photon-number parity.
	That is, $p_{2n}=0$ implies a violation of the first, photon-counting-based criterion in Eq. \eqref{eq:CountingBasedExample}, e.g., when $k+l=2n+1$ is of odd parity and $p_{2n+1}\neq0$.
	Even more interesting is the case of half-integer values, where $k,l\in\frac{1}{2}\mathbb N\setminus\mathbb N$.
	Complementing the integer case, this half-integer scenario is useful for even parity, i.e., $p_{2n+1}=0$.
	For example, the $C$-based criterion in Eq. \eqref{eq:CountingBasedExample} is violated when $k+l=2n$ and $p_{2n}\neq0$.

	Considering the moment-based criterion in Eq. \eqref{eq:MomentBasedExample}, let us take $\mathcal{I}=\{1,2\}$, which yields the witness $\langle{:}\hat n^2{:}\rangle\langle{:}\hat n^{4}{:}\rangle-\langle{:}\hat n^3{:}\rangle^2$ as a nonclassicality test, for example, bounding the third-order moment from above for classical light.
	If, however, we take $\mathcal{I}=\{1/2,3/2\}$, the nonclassicality witness takes the form $\langle{:}\hat n{:}\rangle\langle{:}\hat n^3{:}\rangle-\langle{:}\hat n^2{:}\rangle^2$, which bounds third-order term for classical light from below.
	This is a alternative criterion that cannot be derived from a finite number of integer powers of the photon-number operator.
    However, an infinite-order Taylor series may overcome this limitation;
    but such a series is not accessible with a limited resolution, as considered later.

    It is worth remarking that the above examples for photoelectric measurements can be derived using alternative methods, too \cite{K96a}.
    However, our approach ensures the completeness of the set of matrices that can be constructed and analyzed through choices of the index set $\mathcal I$, rendering it possible to formulate our main findings in Secs. \ref{Sec:ClickCounts} and \ref{Sec:MultiClickCounts}.
    
\subsection{Example}

	As seminal examples of states with different photon-number parities, we investigate the verification of nonclassicality for the superpositions of two coherent states, specifically even and odd cat states \cite{DMM74},
	\begin{equation}
		\label{eq:CatState}
		|\alpha_\pm\rangle
		=
		\frac{
			|\alpha\rangle
			\pm
			|-\alpha\rangle
		}{\sqrt{2\left(
			1\pm e^{-2|\alpha|^2}
		\right)}},
	\end{equation}
	where $\alpha\in\mathbb C$ is the coherent amplitude.
	The even states ($+$) have an even photon-number parity, and odd states ($-$) feature an odd parity.
    Note that for small amplitudes, $|\alpha|\ll1$, these states approximate rather generic states with even and odd photon-number distributions, i.e., $|\alpha_+\rangle\approx x_0|0\rangle+x_2|2\rangle$ and $|\alpha_-\rangle\approx x_1|1\rangle+x_3|x_3\rangle$, for $x_2,x_3\in\mathbb C$ determined via $\alpha$ and $x_0,x_1\in\mathbb C$ accounting for normalization and a global phase.

\begin{figure}
	\includegraphics[width=.49\columnwidth]{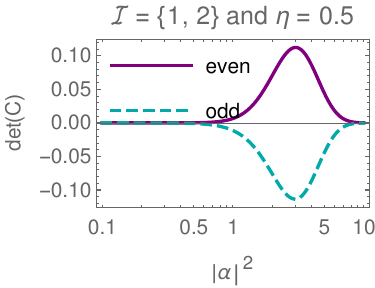}
	\hfill
	\includegraphics[width=.49\columnwidth]{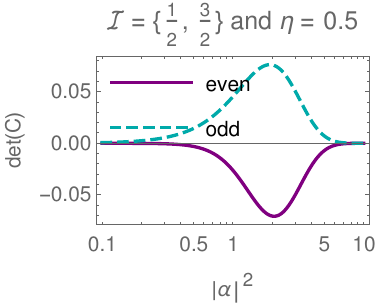}
	\\
	\includegraphics[width=.49\columnwidth]{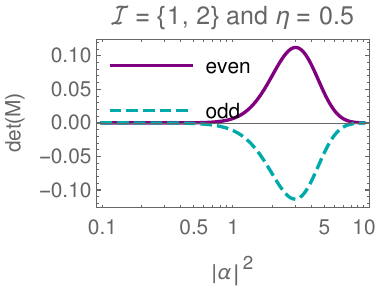}
	\hfill
	\includegraphics[width=.49\columnwidth]{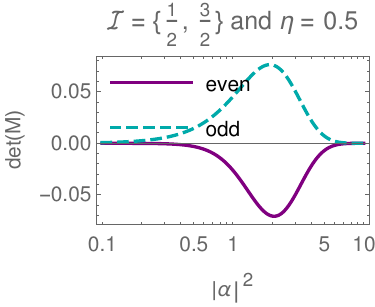}
	\caption{%
		Nonclassicality witnessing through negativities based on the indicated index sets $\mathcal{I}$ with integers (left column) and half-integers (right column) for even (solid) and odd (dashed) cat states $|\alpha_\pm\rangle$.
		For a detection loss of $1-\eta=50\%$, criteria based on the photocounts [top row, cf. Eq. \eqref{eq:CountingBasedExample}] and normally ordered moments [bottom row, cf. Eq. \eqref{eq:MomentBasedExample}] are shown as a function of $|\alpha|^2$ over several orders of magnitude on a logarithmic scale.
		Even-parity cat states are sensitive to the half-integer-based test;
		odd-parity nonclassicality is verified through integer-based criteria.
		Note that the $C$-based and $M$-based nonclassicality probes behave rather similarly for the states in Eq. \eqref{eq:CatState}.
	}\label{fig:photocounting}
\end{figure}

	For determining probabilities and general expectation values with even and odd cat states, we can take the identity
	\begin{equation}
		\label{eq:HelperIdentity}
	\begin{aligned}
		\langle \alpha_\pm|{:}h(\hat n){:}|\alpha_\pm\rangle
		={}&
		\frac{
			h(|\alpha|^2)\pm h(-|\alpha|^2) e^{-2|\alpha|^2}
		}{
			1\pm e^{-2|\alpha|^2}
		}
	\end{aligned}
	\end{equation}
	for an arbitrary function $h$.
	For instance, for $n$ photocounts with a quantum efficiency $\eta$, we have $h(x)=e^{-\eta x}(\eta x)^n/n!$ from the definition in Eq. \eqref{eq:ClickCountsMatrix}.
    Meanwhile, for $m$\textsuperscript{th}-order moments with the same efficiency for Eq. \eqref{eq:MomentMatrix}, we have $h(x)=(\eta x)^m$, allowing us to construct the corresponding matrices $C$ and $M$.
	See Fig. \ref{fig:photocounting} for the distinct criteria obtained from whole and half integer sets $\mathcal I$ that prove nonclassicality for different parities of the even and odd coherent states in Eq. \eqref{eq:CatState}.

	As a final remark, one can replace $\hat n\mapsto \Gamma(\hat n)=\hat \Gamma$ to include general detector response functions $\Gamma$ that account for losses, dark counts, and other detector imperfections \cite{KK64}.
	The most important case is the introduction of losses (cf. Fig. \ref{fig:photocounting}), where $\hat \Gamma=\eta\hat n$ for a non-unit and non-vanishing quantum efficiency, $0< \eta< 1$.

\subsection{Outlook on the remainder of this work}

	Thus far, we considered photoelectric detection with a full---but possibly attenuated \cite{KK64}---photon-number resolution.
	However, a main practical concern is how one can identify nonclassicality from measured statistics and reconstructed moments when such a model is not applicable, mainly due to a limited resolution.
	Thus, in the remainder of this work, we apply the principles introduced above to state-of-the-art, multiplexing-based optical detection schemes.
	Again, special emphasis is put on parity consideration and the role of integer-based and half-integer-based nonclassicality criteria.

	We are going to consider multiplexing of individual on-off detectors---yielding a quantum version of a binomial, rather than Poissonian statistics---in Sec. \ref{Sec:ClickCounts} and individual detectors with a limited intrinsic photon-number resolution---described by multinomial-based statistics---in Sec. \ref{Sec:MultiClickCounts}.
	We begin with the single-mode scenario and later introduce generalizations to the multimode scenario, Sec. \ref{Sec:Multimode}.
	Carefully choosing $\hat f$ and index sets $\mathcal{I}$ in such scenarios is going to render it possible to generalize the treatment from the photocounting case to broad families of multiplexed click-counting scenarios.

\section{Click-counting with on-off detectors}
\label{Sec:ClickCounts}

\begin{figure}
	\includegraphics[width=.8\columnwidth]{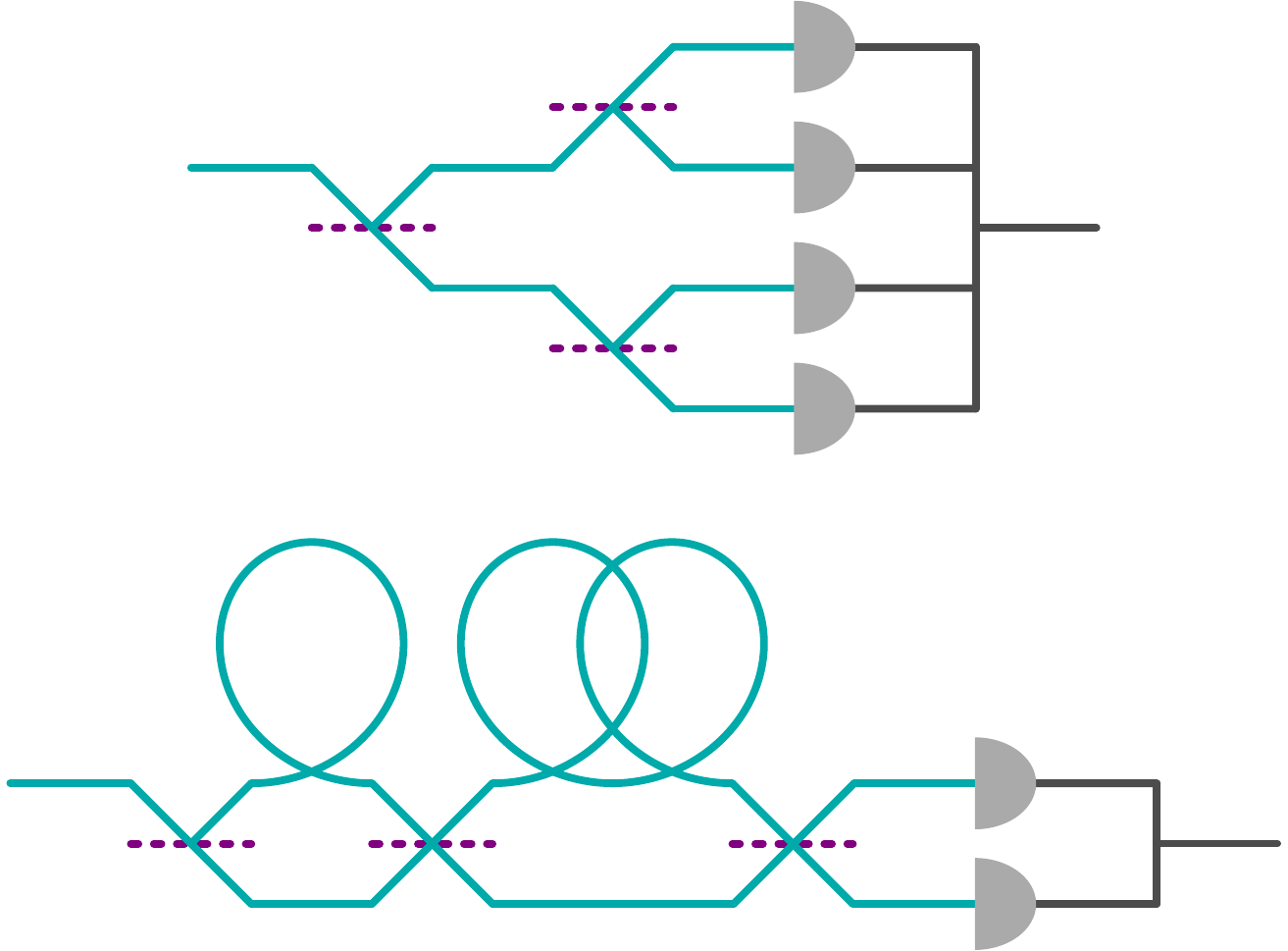}
	\caption{%
		Spatial (top) and time-bin (bottom) multiplexing schemes.
		Light enters from the left and is distributed via $50/50$ beam splitters (dashed lines).
		In time-bin multiplexing (bottom), a delay loop separates pulses in time, while spatial multiplexing uniformly distributes the light across different paths.
		Individual detectors at the end record ``light on'' and ``light off'' information as so-called click and no-click events, respectively.
		The total number $k$ of clicks is recorded.
		The depicted spatial multiplexing (top) uniformly spreads the light into $N=4$ paths, each measured with one on-off detector.
		The shown time-bin multiplexing uniformly distributes light into four time bins for the top and bottom path, resulting in up to $N=8$ total click events, using two on-off detectors only.
		Note that combinations of the two schemes are possible, and that other schemes use uniform illumination of detector arrays, consisting of $N$ diodes \cite{BDFL09,KASVSH17}.
		Alternative schemes, including deviations from a uniform splitting, have been studied, too \cite{LFPR16}.
	}\label{fig:Multiplexing}
\end{figure}

	Now, we consider optical multiplexing \cite{PTKJ96,ASSBW03,KW03,RHHPH03}.
	Two frequently used multiplexed detector configurations are shown in Fig. \ref{fig:Multiplexing}.
	Therein, incident light is uniformly distributed across $N$ spatial and temporal modes.
	Each mode can be assigned a click or no-click from identical on-off detectors, such as APDs in Geiger mode and SNSPDs.
	This results in $k$ total click events, where $0\leq k\leq N$.

	Essential insights are that a click does not equate to a single photon and that the resolution is limited to $N$ clicks, saturating for high intensities.
	This holds true even if the on-off detectors themselves are noiseless and lossless.
    This thus requires a deviation from Poisson-based statistical analyses \cite{SVA12b,SVA13,SECMRKNLGWAV17}.
	In the spirit of the preliminary discussion for photon-number-based nonclassicality witnesses, we now generalize our earlier observations to multiplexed on-off detectors.

\subsection{Nonclassicality criteria in terms of click counts}

	It was shown that the click-counting distribution $c_k$ \cite{SVA12b} for $N$ on-off detectors takes the form of a quantum version of a binomial distribution,
	\begin{equation}
		\label{eq:ClickCounting}
		c_k
		=
		\left\langle{:}
			\binom{N}{k}
			\left(e^{-\hat\Gamma}\right)^{N-k}
			\left(\hat 1-e^{-\hat\Gamma}\right)^{k}
		{:}\right\rangle,
	\end{equation}
	where $\hat \Gamma=\Gamma(\hat n/N)$ is the previously discussed detector response function, additionally accounting for the uniform distribution of the incident photons onto $N$ bins in the argument.
	Remarkably, deviations from uniformity can be additionally considered  \cite{LFPR16}, resulting in a mixed Poisson-binomial model.

	In the same manner outlined for the Poisson-type photoelectric measurements, Eq. \eqref{eq:GeneralWitness}, we can now construct nonclassicality witnesses  \cite{SVA13}.
	Firstly, we take
	\begin{equation}
		\hat f
		=
		\sum_{k\in \mathcal{I}}
		f_k
		e^{-(N/2-k)\hat\Gamma}
		\left(\hat 1-e^{-\hat\Gamma}\right)^{k}.
	\end{equation}
	From this, we proceed as done in Sec. \ref{Sec:PhotonCounts} and can immediately deduce the following constraint for classical light in terms of the matrix of click counts $C$:
	\begin{equation}
		\label{eq:WitnessClickCounts}
		0
		\stackrel{\text{cl.}}{\leq}
		C
		=
		\left[
			\binom{N}{k+l}^{-1} c_{k+l}
		\right]_{k,l\in \mathcal{I}}.
	\end{equation}
	The same reasoning that led us to the half-full-integer discrimination of the index set $\mathcal{I}$ [cf. Eq. \eqref{eq:SelectionIndexSet}] can be applied here, together with the additional constraint $k+l\leq N$ for the click-counting scenario.
	Thus, we have
	\begin{equation}
		\label{eq:OnOffClickIndexSets}
	\begin{aligned}
		\text{either}\quad
		\mathcal{I}
		\subseteq{}&
		\left\{
			0,1,\ldots,
			\left\lfloor
				\frac{N}{2}
			\right\rfloor
		\right\}
		\subset\mathbb N
		\\
		\text{or}\quad
		\mathcal{I}
		\subseteq{}&
		\left\{
			\frac{1}{2},\frac{3}{2},\ldots,
			\left\lceil
				\frac{N}{2}
			\right\rceil
			-\frac{1}{2}
		\right\}\subset\frac{1}{2}\mathbb N\setminus\mathbb N,
	\end{aligned}
	\end{equation}
	ensuring that, for even and odd $N$, all the elements $k,l$ are considered such that $k+l\leq N$ and $k+l\in\mathbb N$ hold true, importantly including the case $k=l$.

	As an explicit example, we take $N=4$ and $\mathcal{I}=\{0,1,2\}$ consisting of whole numbers, resulting in
	\begin{equation}
		C
		=
		\begin{bmatrix}
			c_0/1 & c_1/4 & c_2/6
			\\
			c_1/4 & c_2/6 & c_3/4
			\\
			c_2/6 & c_3/4 & c_4/1
		\end{bmatrix}.
	\end{equation}
	For a classical state, Eq. \eqref{eq:WitnessClickCounts} states that $C$ is positive semidefinite, meaning that all eigenvalues are non-negative, which is equivalent to
	\begin{equation}
		\label{eq:MinEigCounting}
		0
		\stackrel{\text{cl.}}{\leq}\min\mathrm{eig}(C).
	\end{equation}
	For a half-integer index set, e.g., $\mathcal{I}=\{1/2,3/2\}$, also taking $N=4$, we obtain the click-counting matrix
	\begin{equation}
		C
		=
		\begin{bmatrix}
			c_1/1 & c_2/3
			\\
			c_2/3 & c_3/1
		\end{bmatrix},
	\end{equation}
	which also obeys the relation in Eq. \eqref{eq:MinEigCounting} for classical states of light.
	Again, we see that independent nonclassicality criteria emerge when whole and half integers are used to construct quantumness tests for the experimentally measured click-counting statistics.

\begin{figure}
	\includegraphics[width=.8\columnwidth]{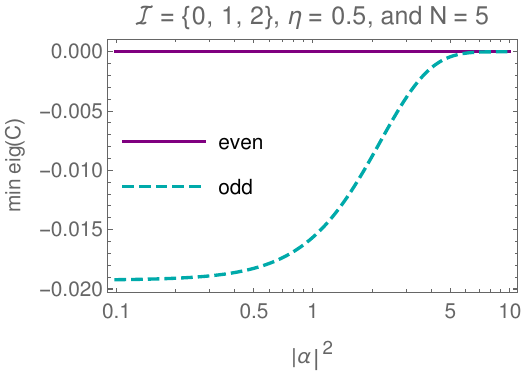}
	\\[2ex]
	\includegraphics[width=.8\columnwidth]{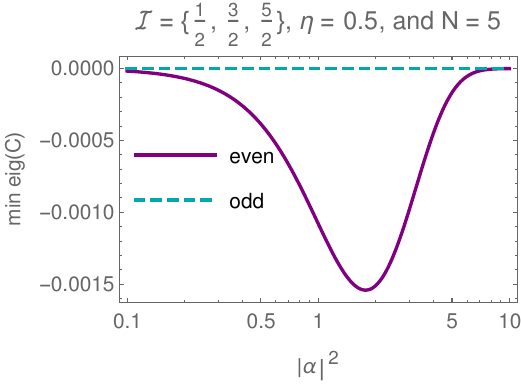}
	\caption{%
		Nonclassicality witnessing through negative eigenvalues in the click-counting matrix $C$, Eq. \eqref{eq:WitnessClickCounts}.
		Each on-off detector has an efficiency of only $50\%$, and we make the unconventional choice of $N=5$ to highlight the applicability of our method even in the case of an odd number of detection bins.
        (For even $N$s, the approach works analogously.)
		The index set with whole numbers (top plot) verifies the nonclassicality of the odd coherent state $|\alpha_{-}\rangle$ but is insensitive to the even coherent states $|\alpha_{+}\rangle$.
		Conversely, the index set with half integers (bottom plot) is sensitive to the opposite parity.
	}\label{fig:ClickCountingMatrix}
\end{figure}

	In Fig. \ref{fig:ClickCountingMatrix}, we apply the criteria defined by Eq. \eqref{eq:WitnessClickCounts} to even and odd cat states from Eq. \eqref{eq:CatState}, using the response function $\hat\Gamma=\Gamma(\hat n/N)=\eta\hat n/N$ to account for losses.
	Taking the click-counting statistics in Eq. \eqref{eq:ClickCounting}, together with the identity in Eq. \eqref{eq:HelperIdentity}, we can construct the matrix $C$ of click counts and evaluate its eigenvalues for different sets $\mathcal{I}$.
	Again, we see that the integer and half-integer index sets enable us to certify the nonclassicality of states with different parities.

\subsection{Matrix of click moments}

	Click-based, normally ordered moments can be directly determined from click-counting data as \cite{SVA13}
	\begin{equation}
		\left\langle{:}
			\hat\pi^m
		{:}\right\rangle
		=\sum_{k=m}^{N}\frac{\binom{k}{m}}{\binom{N}{m}} c_k,
		\quad\text{with}
		\quad
		\hat\pi=\hat 1-\exp(-\hat\Gamma)
	\end{equation}
	being the click analog to the photon-number operator.
	For moments, we can again follow an analogous construction as done in Sec. \ref{Sec:PhotonCounts}, yielding
	\begin{equation}
		\label{eq:WitnessClickMoments}
		\hat f=\sum_{k\in \mathcal{I}} f_k\hat\pi_k
		\quad\text{and}\quad
		0
		\stackrel{\text{cl.}}{\leq}
		M
		=
		\left[
			\left\langle{:}
				\hat\pi^{k+l}
			{:}\right\rangle
		\right]_{k,l\in \mathcal{I}},
	\end{equation}
	defining the matrix of click moments $M$ constructed with the half-integer and whole-integer index sets $\mathcal{I}$ discussed in Eq. \eqref{eq:OnOffClickIndexSets}.
	In either case, i.e., half or whole integer sets $\mathcal{I}$, we have a non-negative matrix of click moments for classical light,
	\begin{equation}
		\label{eq:MinEigMoments}
		0
		\stackrel{\text{cl.}}{\leq}
		\min\mathrm{eig}(M).
	\end{equation}
	For the example of $N=4$ spatial and temporal detection bins,
	we can maximally select $\mathcal{I}=\{0,1,2\}$ and $\mathcal{I}=\{1/2,3/2\}$, resulting in
	\begin{equation}
	\begin{aligned}
		M
		=
		\begin{bmatrix}
			1
			&
			\left\langle{:}
				\hat\pi
			{:}\right\rangle
			&
			\left\langle{:}
				\hat\pi^2
			{:}\right\rangle
			\\
			\left\langle{:}
				\hat\pi
			{:}\right\rangle
			&
			\left\langle{:}
				\hat\pi^2
			{:}\right\rangle
			&
			\left\langle{:}
				\hat\pi^3
			{:}\right\rangle
			\\
			\left\langle{:}
				\hat\pi^2
			{:}\right\rangle
			&
			\left\langle{:}
				\hat\pi^3
			{:}\right\rangle
			&
			\left\langle{:}
				\hat\pi^4
			{:}\right\rangle
		\end{bmatrix}
		\\
		\text{and}\quad
		M
		=
		\begin{bmatrix}
			\left\langle{:}
				\hat\pi
			{:}\right\rangle
			&
			\left\langle{:}
				\hat\pi^2
			{:}\right\rangle
			\\
			\left\langle{:}
				\hat\pi^2
			{:}\right\rangle
			&
			\left\langle{:}
				\hat\pi^3
			{:}\right\rangle
		\end{bmatrix},
	\end{aligned}
	\end{equation}
	respectively.
	The first line results in an up to fourth-order nonclassicality criterion while the second line is up to third order.
	The maximal order for integer-based and half-integer-based criteria is limited by the click-detector resolution $N$ as given in Eq. \eqref{eq:OnOffClickIndexSets}.

\begin{figure}
	\includegraphics[width=.8\columnwidth]{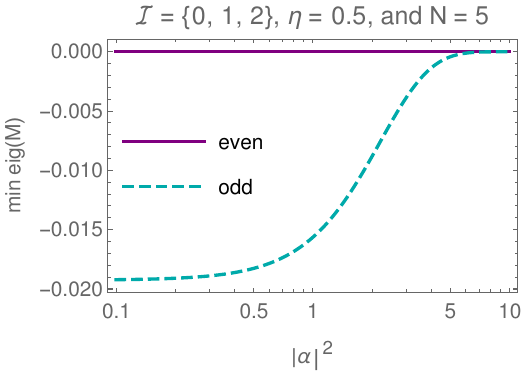}
	\\[2ex]
	\includegraphics[width=.8\columnwidth]{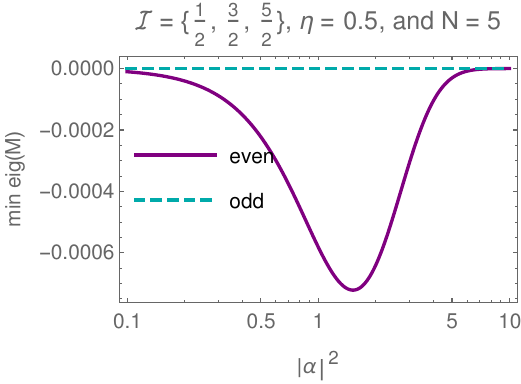}
	\caption{%
		Nonclassicality witnessing through negative eigenvalues in the matrix $M$ of click moments, Eq. \eqref{eq:MinEigMoments}.
		The settings of the detection system are the same as in Fig. \ref{fig:ClickCountingMatrix}.
		Note that, despite the different physical and statistical meaning of the matrix $C$ in Fig. \ref{fig:ClickCountingMatrix} and the matrix $M$ here, the plots here are technically identical because $C$ can be mapped to $M$ without altering signs of eigenvalues, using generating functions \cite{SVA13}.
	}\label{fig:ClickMomentMatrix}
\end{figure}

	In Fig. \ref{fig:ClickMomentMatrix}, we apply the criteria that pertain to the minimal eigenvalues of matrices of moments $M$ in Eq. \eqref{eq:MinEigMoments} for different types of index sets $\mathcal{I}$.
	Like before, we use Eq. \eqref{eq:HelperIdentity} to obtain exact expressions for the entries of the matrix of moments.
	Our previous observation stated that, beyond the well-known integer-valued $\mathcal{I}$ for odd parity, half-integer sets are needed to uncover the nonclassical features of even-parity states.

\subsection{Additional statistical properties}

	In addition to the general framework formulated above, it is convenient to discuss additional interpretations of the click moments and their physical meaning.
	This extra discussion allows us to contextualize our previously presented findings.

\subsubsection{Counting-based criteria}

	The Klyshko criterion \cite{K96} applies to the photon-number statistics and relates photoelectric counts to assess a state's nonclassicality.
	Equation \eqref{eq:WitnessClickCounts} allows us to achieve the same for multiplexed on-off detectors.
	For example, we can take whole integers $\mathcal I=\{0,1\}$,
	\begin{equation}
	\begin{aligned}
		0
		\stackrel{\text{cl.}}{\leq}
		{}&
		\det
		\begin{bmatrix}
			\binom{N}{0}^{-1} c_{0}
			&
			\binom{N}{1}^{-1} c_{1}
			\\
			\binom{N}{1}^{-1} c_{1}
			&
			\binom{N}{2}^{-1} c_{2}
		\end{bmatrix}
		\\
		\Leftrightarrow\quad
		\frac{1}{2}\left(1-\frac{1}{N}\right)
		\stackrel{\text{cl.}}{\leq}
		{}&
		\frac{c_0c_2}{c_1^2},
	\end{aligned}
	\end{equation}
	which allows us to relate coincidence clicks $c_2$ with single click counts $c_1$, without the need for challenging inversions of the click distribution into a photon-number distribution \cite{KSADSBSS24}.
	For half-integers, e.g., $\mathcal I=\{1/2,3/2\}$, we can write analogously
	\begin{equation}
	\begin{aligned}
		0
		\stackrel{\text{cl.}}{\leq}
		{}&
		\det
		\begin{bmatrix}
			\binom{N}{1}^{-1} c_{1}
			&
			\binom{N}{2}^{-1} c_{2}
			\\
			\binom{N}{2}^{-1} c_{2}
			&
			\binom{N}{3}^{-1} c_{3}
		\end{bmatrix}
		\\
		\Leftrightarrow\quad
		\frac{2}{3}
		\left(1-\frac{1}{N-1}\right)
		\stackrel{\text{cl.}}{\leq}
		{}&
		\frac{c_1 c_3}{c_2^2},
	\end{aligned}
	\end{equation}
	which enables us to relates three-fold clicks $c_3$ with coincidence and single counts.
	When violating the inequalities formulated above, we directly witness the nonclassicality of the probed quantum state of light.

\subsubsection{Relation to correlation functions}

	Since Glauber's pioneering work \cite{G63a}, correlation functions have played a critical role in assessing an optical state's quantumness.
	As mentioned above, the operator $\hat\pi=\hat 1-e^{-\hat\Gamma}$ relates to click statistics $c_k$ in the same manner as the number operator relates to the photon-number distribution $p_n$.
	Thus, analogously to correlation functions, given by $\left\langle{:}\hat n^m{:}\right\rangle/\left\langle{:}\hat n{:}\right\rangle^m$, we can define the click-counting counterpart as the normalized $m$\textsuperscript{th}-order click-correlation function
	\begin{equation}
		g^{(m)}
		=
		\frac{
			\left\langle{:}\hat\pi^{m}{:}\right\rangle
		}{
			\left\langle{:}\hat\pi{:}\right\rangle^{m}
		}.
	\end{equation}

	In terms of the matrix of click moments, Eq. \eqref{eq:WitnessClickMoments}, $g^{(k+l)}$ requires that we multiply the matrix $M$ from the left and right with the positive diagonal matrix
	\begin{equation}
		D
		=
		\mathrm{diag}\left[
			\left\langle{:}\hat\pi{:}\right\rangle^{-m}
		\right]_{m\in \mathcal{I}}=D^\dag.
	\end{equation}
	Now, the following property applies:
	\begin{equation}
		0
		\stackrel{\text{cl.}}{\leq}
		M
		\quad\Leftrightarrow\quad
		0
		\stackrel{\text{cl.}}{\leq}
		DMD^\dag.
	\end{equation}
	The entries of $DMD^\dag$ are the sought click-based correlation functions.
	For example, we find
	\begin{equation}
		0
		\stackrel{\text{cl.}}{\leq}
		\det
		\begin{bmatrix}
			1 & g^{(1)} \\ g^{(1)} & g^{(2)}
		\end{bmatrix}
		=g^{(2)}-1,
	\end{equation}
	for $\mathcal{I}=\{0,1\}$ and using $g^{(1)}=1$, and
	\begin{equation}
		0
		\stackrel{\text{cl.}}{\leq}
		\det
		\begin{bmatrix}
			1 & g^{(2)} \\ g^{(2)} & g^{(3)}
		\end{bmatrix}
		=g^{(3)}-\left(g^{(2)}\right)^2,
	\end{equation}
	for $\mathcal{I}=\{1/2,3/2\}$, bounding the third-order correlation function.

\subsubsection{Statistical quantifiers}

	The third analysis pertains to statistical quantities that describe a probability distribution.
	For instance, from previous works on moments of click-counting statistics \cite{SVA12,SVA13}, it is known that
	\begin{subequations}
	\begin{align}
		\left\langle{:}
			\hat\pi
		{:}\right\rangle
		={}&
		\frac{\mathsf{E}[k]}{N},
		\\
		\left\langle{:}
			\hat\pi^2
		{:}\right\rangle
		={}&
		\frac{\mathsf{E}[k(k-1)]}{N(N-1)},
		\\
		\left\langle{:}
			\hat\pi^3
		{:}\right\rangle
		={}&
		\frac{\mathsf{E}[k(k-1)(k-2)]}{N(N-1)(N-2)},
	\end{align}
	\end{subequations}
	where $\mathsf{E}[f(k)]=\sum_{k=0}^N c_k f(k)$ denotes the expected value for a function $f$.
	In addition, the first three moments are related to important statistical quantities
	\begin{subequations}
	\begin{align}
		\mu
		={}&
		\mathsf{E}\left[k\right],
		\\
		\sigma^2
		={}&
		\mathsf{E}\left[\left(k-\mu\right)^2\right],
		\\
		\gamma_3
		={}&
		\mathsf{E}\left[\left(\frac{k-\mu}{\sigma}\right)^3\right],
	\end{align}
	\end{subequations}
	being the mean value $\mu$, the variance $\sigma^2$, and the skewness $\gamma_3$, respectively.
	Solving for the different orders $\mathsf{E}[k^m]$ ($m\in\{1,2,3\}$) and substituting the results into the previous relations, we obtain
	\begin{subequations}
		\label{eq:ThreeClickMoments}
	\begin{align}
		\left\langle{:}
			\hat\pi
		{:}\right\rangle
		={}&
		\frac{
			\mu
		}{
			N
		},
		\\
		\left\langle{:}
			\hat\pi^2
		{:}\right\rangle
		={}&
		\frac{
			\sigma^2+\mu^2-\mu
		}{
			N(N-1)
		},
		\\
		\left\langle{:}
			\hat\pi^3
		{:}\right\rangle
		={}&
		\frac{
			\sigma^3\gamma_3
			+3\mu\sigma^2
			+\mu^3
			-3\left(\sigma^2+\mu^2\right)
			+2\mu
		}{
			N(N-1)(N-2)
		}.
	\end{align}
	\end{subequations}

	In our example with integers, $\mathcal{I}=\{0,1\}$, we can take Eq. \eqref{eq:WitnessClickMoments}, compute its determinant, and recast it into the following form:
	\begin{equation}
	\begin{aligned}
		{}&
		\det
		\begin{bmatrix}
			1 & \langle{:}\hat\pi{:}\rangle
			\\
			\langle{:}\hat\pi{:}\rangle & \langle{:}\hat\pi^2{:}\rangle
		\end{bmatrix}
		=
		\left\langle{:}
			\hat\pi^2
		{:}\right\rangle
		-
		\left\langle{:}
			\hat\pi
		{:}\right\rangle^2
		\\
		={}&
		\frac{
			\mu(N-\mu)
		}{
			N^2(N-1)
		}
		\left(
			\frac{N\sigma^2}{\mu(N-\mu)}
			-1
		\right).
	\end{aligned}
	\end{equation}
	This case includes in parenthesis the binomial parameter $Q_{B}=N\sigma^2/[\mu(N-\mu)]-1$ that was constructed \cite{SVA13} and experimentally applied \cite{BDJDBW13} for assessing nonclassicality via the variance $\sigma^2$ and mean $\mu$ of the measured click-counting statistic.
	It replaces the seminal Mandel $Q$ parameter \cite{M79}, which yields accurate results only in case of photon-counting.
	Beyond this known parameter, the example of the half-integer set $\mathcal I=\{1/2,3/2\}$ allows us to formulate an analogous, yet previously unknown relation for the skewness $\gamma_3$.
	Applying the above identities, we find after some algebra
	\begin{equation}
		\label{eq:ClickSkewnessCriterion}
	\begin{aligned}
		{}&
		\det
		\begin{bmatrix}
			\langle{:}\hat\pi{:}\rangle & \langle{:}\hat\pi^2{:}\rangle
			\\
			\langle{:}\hat\pi^2{:}\rangle & \langle{:}\hat\pi^3{:}\rangle
		\end{bmatrix}
		=
		\left\langle{:}
			\hat\pi
		{:}\right\rangle
		\left\langle{:}
			\hat\pi^3
		{:}\right\rangle
		-
		\left\langle{:}
			\hat\pi^2
		{:}\right\rangle^2
		\\
		={}&
		\langle{:} \hat\pi {:}\rangle
		\langle{:} (\Delta\hat\pi)^3 {:}\rangle
		+
		\langle{:} \hat\pi {:}\rangle^2
		\langle{:} (\Delta\hat\pi)^2 {:}\rangle
		-
		\langle{:} (\Delta\hat\pi)^2 {:}\rangle^2,
	\end{aligned}
	\end{equation}
	where $\Delta\hat\pi=\hat\pi-\langle{:}\hat\pi{:}\rangle$, $\langle{:} \hat\pi {:}\rangle=\mu/N$, $\bar\mu=N-\mu$,
	\begin{align}
		\langle{:} (\Delta\hat\pi)^2 {:}\rangle
		={}&
		\frac{
			N\sigma^2-\mu\bar\mu
		}{
			N^2(N-1)
		},
		\quad\text{and}
		\\ \nonumber
		\langle{:} (\Delta\hat\pi)^3 {:}\rangle
		={}&
		\frac{
			N^2\sigma^3\gamma_3
			+2\mu\bar\mu(\bar\mu-\mu)
			-3N\sigma^2(\bar\mu-\mu)
		}{
			N^3(N-1)(N-2)
		}.
	\end{align}
	For light with a classical binomial click statistics, the variance and skewness can be expressed as $N\sigma^2=\mu\bar\mu$ and $N^2\sigma^3\gamma_3=N\sigma^2(\bar\mu-\mu)=\bar\mu\mu(\bar\mu-\mu)$, implying $\langle{:} (\Delta\hat\pi)^2 {:}\rangle=0$ and $\langle{:} (\Delta\hat\pi)^3 {:}\rangle=0$.
	For nonclassical light, however, the expression in Eq. \eqref{eq:ClickSkewnessCriterion} can become negative, thus presenting an additional and clearly nontrivial nonclassicality witness.

\section{Multiplexing with some intrinsic photon-number resolution}
\label{Sec:MultiClickCounts}

	Beyond on-off detectors, we can replace the individual detectors in Fig. \ref{fig:Multiplexing} with detectors that have a partial intrinsic photon-number resolution, such as TESs \cite{LMN08} and as recently demonstrated for SNSPDs \cite{SLHSSBSB24,SKGCS23}.
	With this intrinsic resolution, we have, in addition to the no-click event, up to $K$ distinct types of click events.
	Considering the outcomes from all bins, we have the number $N_j$ of bins that count the $j$th intrinsic event type, for $j\in\{0,\ldots, K\}$.
	Summing over the $K+1$ outcomes, we get $N=N_0+\cdots+N_K$, i.e., the number of available bins.
	The advantaged gained by multiplexing such detectors with partial photon-number resolution has been experimentally verified, for example, in Refs. \cite{SCEMRKNLGVAW17,SECMRKNLGWAV17,SPBTEWLNLGVASW20}.

\subsection{Nonclassicality witnesses}

	Suppose each individual detector is described by the measurement operator $\hat\pi_j$ for the $j$th outcome.
	Then, it has been shown that the counting statistics of the full, multiplexed system is described by a multinomial distribution \cite{SCEMRKNLGVAW17,SECMRKNLGWAV17},
	\begin{equation}
		\label{eq:MultinomialDistribution}
		c_{N_0,\ldots,N_K}
		=
		\left\langle{:}
			\binom{N}{N_0,\ldots,N_K}
			\hat\pi_0^{N_0}
			\cdots
			\hat\pi_K^{N_K}
		{:}\right\rangle,
	\end{equation}
	with the multinomial coefficient $\binom{N}{N_0,\ldots,N_K}=N!/(N_0!\cdots N_K!)$ and $N=N_0+\cdots+N_K$.
	Note that the special case $K=1$ yields the previous on-off detection scenario with $N_0=N-k$ and $N_1=k$.
	Furthermore, if the individual detection outcomes directly relate to the photon number, we have
	\begin{equation}
	\begin{aligned}
		\label{eq:fMultinomial}
		\hat\pi_j
		={}&
		{:}\frac{\hat\Gamma^j}{j!}e^{-\hat\Gamma}{:},
		\text{ for $0\leq j< K$},
		\\
		\text{and }
		\hat\pi_K
		={}&
		\hat 1-\left(\hat\pi_0+\cdots+\hat\pi_{K-1}\right),
	\end{aligned}
	\end{equation}
	with the response function $\hat\Gamma=\Gamma(\hat n/N)$.
    That is, the click events $0\leq j<K$ correspond to actual photon numbers in this scenario, and the $K$\textsuperscript{th} (i.e., last) bin collects all events which are not clearly assigned to the photon numbers from zero to $K-1$.
	Again, we obtain the previously discussed $\hat\pi_0={:}e^{-\hat\Gamma}{:}$ and $\hat\pi_1=\hat1-{:}e^{-\hat\Gamma}{:}$ for on-off detectors if $K=1$.

	For the witnessing, cf. Secs. \ref{Sec:PhotonCounts} and \ref{Sec:ClickCounts}, we can choose
	\begin{equation}
        \label{eq:MultinomialMat}
		\hat f
		=
		\sum_{(N_0,\ldots,N_K)\in\mathcal I}
		f_{N_0,\ldots,N_K}\hat \pi_0^{N_0} \cdots \hat\pi_K^{N_K}
		,
	\end{equation}
	for a generic index set $\mathcal I$ such that $N/2=N_0+\cdots+N_K$ for all multi-indices $(N_0,\ldots,N_K)\in\mathcal I$.
	As a classical constraint, this yields
	\begin{equation}
	\begin{aligned}
		0
		\stackrel{\text{cl.}}{\leq}
		\sum_{(N_0,\ldots,N_K),(N'_0,\ldots,N'_K)\in\mathcal I}
		{}&
		f_{N_0,\ldots,N_K}^\ast f_{N'_0,\ldots,N'_K}
		\\
		{}&\times
		\frac{c_{N_0+N'_0,\ldots,N_K+N'_K}}{\binom{N}{N_0+N'_0,\ldots,N_K+N'_K}},
	\end{aligned}
	\end{equation}
	allowing us to define the multi-index counting matrix
	\begin{equation}
		\label{eq:MultinomialMatrixCounts}
		C=\left[
			\frac{c_{N_0+N'_0,\ldots,N_K+N'_K}}{\binom{N}{N_0+N'_0,\ldots,N_K+N'_K}}
		\right]_{(N_0,\ldots,N_K),(N'_0,\ldots,N'_K)\in\mathcal I},
	\end{equation}
	which has to be positive semidefinite for classical light, as proven in the previous sections.
	We emphasize that the multi-index $(N_0,\ldots,N_K)$ refers to the $K+1$ different outcomes of the individual detectors that probe a single-mode incident light field;
	this is different from the multimode discussion in the following Sec. \ref{Sec:Multimode}.

	In a similar manner as described above, we take $\hat f$ as given in Eq. \eqref{eq:MultinomialMat}.
	However, we now interpret the multi-index as the corresponding moments, resulting in the matrix of moments
	\begin{equation}
		M=\left[
			\langle{:}
				\hat\pi_0^{N_0+N'_0}
				\cdots
				\hat\pi_K^{N_K+N'_K}
			\rangle{:}
		\right]_{(N_0,\ldots,N_K),(N'_0,\ldots,N'_K)\in\mathcal I}.
	\end{equation}
	Here, the elements of the multi-index set $\mathcal I$ can be chosen as $N_0+\cdots+N_K\leq N/2$, which is less restricted than for $C$.
	Still, we know that the matrix $M$ of normally ordered click moments is positive semidefinite for classical light, and negativities uncover nonclassical light.

\begin{figure*}
	\includegraphics[width=.45\textwidth]{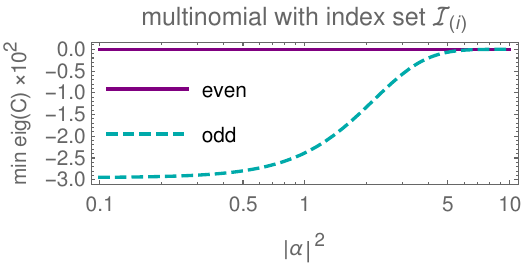}
	\hfill
	\includegraphics[width=.45\textwidth]{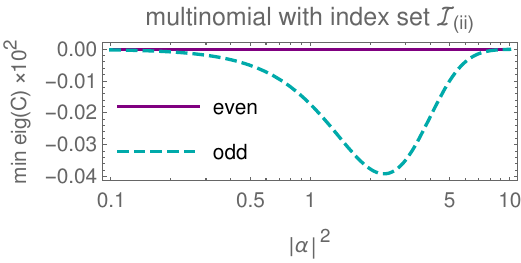}
	\\[2ex]
	\includegraphics[width=.45\textwidth]{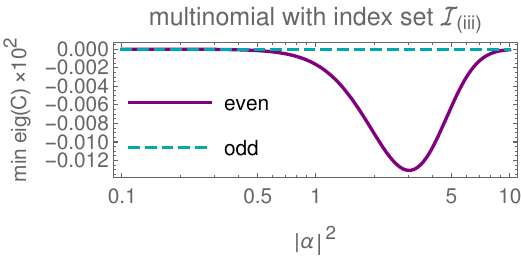}
	\hfill
	\includegraphics[width=.45\textwidth]{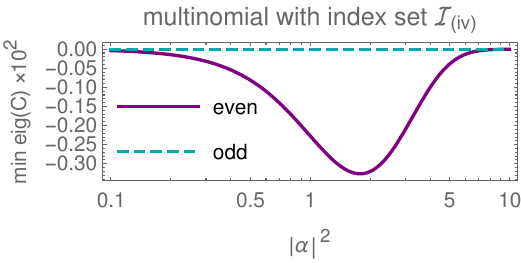}
	\caption{%
		Witnessing nonclassicality via a negative minimal eigenvalue of the multinomial click-counting matrix $C$ in Eq. \eqref{eq:MultinomialMatrixCounts}, using the four possible index sets in Eq.
		\eqref{eq:MultinomialIndexSets}.
		The elements of these index sets $(N_0,N_1,N_2)$ pertain to the number of detectors $N_j$ with no photons detected ($j=0$), one photon detected ($j=1$), and at least two-photons detected ($j=2$), assuming a detection efficiency of $\eta=50\%$.
		The top-left plot certifies the nonclassicality of the odd cat state using integer-based indices $(N_0,N_1,N_2)\in\mathbb N\times\mathbb N\times\mathbb N$.
		The top-right plot achieves this too, but with a generally reduced value of negativity, using $(N_0,N_1,N_2)\in\left(\frac{1}{2}\mathbb N\setminus\mathbb N\right)\times \mathbb N\times\left(\frac{1}{2}\mathbb N\setminus\mathbb N\right)$.
		The bottom-left plot certifies the nonclassicality of the even state via $(N_0,N_1,N_2)\in \mathbb N\times\left(\frac{1}{2}\mathbb N\setminus\mathbb N\right)\times\left(\frac{1}{2}\mathbb N\setminus\mathbb N\right)$.
		The bottom-right plot does the same, but with generally higher negativities when using $(N_0,N_1,N_2)\in\left(\frac{1}{2}\mathbb N\setminus\mathbb N\right)\times\left(\frac{1}{2}\mathbb N\setminus\mathbb N\right)\times\mathbb N$.
	}\label{fig:Multinomial}
\end{figure*}

\subsection{Exponentially more criteria}

	The set of multi-indices $\mathcal I$ can include full and half integers for each individual index $N_j$.
	For instance, suppose that we have $\mathcal I\subseteq\mathcal I_0\times\cdots\times \mathcal I_{K}$, such that $N_j\in\mathcal I_j$ and $N_0+\cdots+N_K=N/2$ for all members of $\mathcal I$ that define the elements of $C$---likewise $N_0+\cdots+N_K\leq N/2$ for $M$.
	For each $0\leq j<K$, we can either select full integers or half-integers for the set $\mathcal I_j$ such that $\forall N_j,N'_j\in\mathcal I_j: N_j+N'_j\in\mathbb N$, while the last index ($N_K\in\mathcal I_K$) is chosen such that $N_0+\cdots+N_K=N/2$ holds true for $C$.
	(Again, the latter constraint does not apply to $M$.)
	Even in the more restricted case of $C$, the permissible choices leaves us with $2^K$ sets $\mathcal I$ for constructing $C$ matrices with multi-indices consisting of whole and half integers.
	When at least one of those choices exhibits a negativity, nonclassical light is certified.
	Thus, we here have an exponentially increasing number of criteria that were previously unavailable and go beyond the known option where all $N_j$ are whole numbers.

\subsection{Example}

	As concrete examples, we consider $N=4$ detection bins and a photon-number-resolution limit of $K=2$, giving $K+1=3$ distinct outcomes per individual bin.
	That is, each of the detectors in Fig. \ref{fig:Multiplexing} can distinguish between no photons ($j=0$), one photon ($j=1$), and more than one photon ($j=K=2$), which are multiplexed in two steps, resulting in four bins.
	In this application, the response function is $\hat\Gamma=\eta\hat n/N$ and the individual operators $\hat\pi_j$ that define the counting statistics in Eq. \eqref{eq:MultinomialDistribution} take the form
	\begin{equation}
	\begin{aligned}
		\hat\pi_0={:}e^{-\hat\Gamma}{:},
		\quad
		\hat\pi_1={:}\hat\Gamma e^{-\hat\Gamma}{:},
		\quad\text{and}
		\\
		\hat\pi_2=\hat 1-\hat\pi_0-\hat\pi_1
		={:}\left(e^{\hat \Gamma}-\hat 1-\hat\Gamma\right)e^{-\hat\Gamma}{:}.
	\end{aligned}
	\end{equation}
	The above construction of $C$, Eq. \eqref{eq:MultinomialMatrixCounts}, yields the following $4=2^K$ distinct choices for the index sets:
	\begin{subequations}
		\label{eq:MultinomialIndexSets}
	\begin{align}
		\nonumber
		\mathcal I_\mathrm{(i)}
		={}&
		\{(0,0,2),(1,0,1),(0,1,1),
		\\
		{}&
		(1,1,0),(2,0,0),(0,2,0)\},
		\\
		\mathcal I_\mathrm{(ii)}
		={}&
		\{
			(1/2,0,3/2),(3/2,0,1/2),(1/2,1,1/2)
		\},
		\\
		\mathcal I_\mathrm{(iii)}
		={}&
		\{
			(0,1/2,3/2),(0,3/2,1/2),(1,1/2,1/2)
		\},
		\\
		\mathcal I_\mathrm{(iv)}
		={}&
		\{
			(1/2,1/2,1),(3/2,1/2,0),(1/2,3/2,0)
		\},
	\end{align}
	\end{subequations}
	being all the combinations of integer and half-integer indices $N_0$, $N_1$, and $N_2=N/2-N_0-N_1$.
	Note that we focus on $C$ as the more restrictive part to avoid redundant discussions;
	implementations with $M$ yield eight options ($2^{K+1}$) but function in a manner very similar to what is discussed here.

	Figure \ref{fig:Multinomial} shows the minimal eigenvalues of the four counting matrices, from $C_\mathrm{(i)}$ to $C_\mathrm{(iv)}$, for the four options of the aforementioned index sets.
	The entries of the matrices are obtained for the even and odd parity states in Eq. \eqref{eq:CatState}, again applying Eq. \eqref{eq:HelperIdentity} for obtaining exact expressions.
	The top plots, pertaining to $\mathcal I_\mathrm{(i)}$ and $\mathcal I_\mathrm{(ii)}$, are suited to witness the state $|\alpha_-\rangle$ but unsuited for $|\alpha_+\rangle$.
	Conversely, the bottom plots, for $\mathcal I_\mathrm{(iii)}$ and $\mathcal I_\mathrm{(iv)}$, successfully witness the nonclassicality of the even coherent state but not the odd coherent state.

	The multi-indices $(N_0,N_1,N_2)\in\mathcal I_\mathrm{(i)}$ contain whole integers only.
	The index set $\mathcal I_\mathrm{(ii)}$ consists of half integers for the vacuum ($N_0$) and integers for a single-photon ($N_1$) and two-and-more-photon contributions ($N_2$), making sure that the elements add to $N/2$.
	The other way around, the index set $\mathcal I_\mathrm{(iii)}$ consists of integers for the vacuum ($N_0$) and half integers for the non-zero photon contributions ($N_1$ and $N_2$).
    Meanwhile, half-integers for $N_0$ and $N_1$ can be found in $\mathcal I_\mathrm{(iv)}$.
	Depending on the choice of index set, the plots in Fig. \ref{fig:Multinomial} exhibit a different sensitvity to nonclassicality with respect to the photon-number parity. This demonstrates the versatility of our approach in highlighting different nonclassicality features.

\section{Multimode generalizations}
\label{Sec:Multimode}

	Thus far, we focused on a single mode.
	However, click-counting detection with a large number of modes has been implented in theory (e.g., Ref. \cite{SVA13}) and experiment (e.g., Refs. \cite{SBVHBAS15,TEBSS21,SBDBJDVW16}) for analyzing nonclassical correlations.
	In a multimode scenario, each mode is measured with a click-counting device.
	Since we laid out how to proceed from photon counts, Sec. \ref{Sec:PhotonCounts}, to click counts via either on-off detectors, Sec. \ref{Sec:ClickCounts}, or detectors with intrinsic resolutions, Sec. \ref{Sec:MultiClickCounts}, we solely focus on the generalization of the full integer and half-integer classification.
	Thus, we are going to discuss photon-number-based nonclassicality criteria in multimode systems and explore correlation criteria that are formulated via integer and half-integer index sets.

\subsection{Multimode criteria}

	Suppose we have $\mu$ modes, then we can formulate $\mu$-mode classical constraints for the multimode counting matrices $C$ and matrices $M$ of normally ordered moments according to the principles introduced previously,
	\begin{equation}
	\begin{aligned}
		0\stackrel{\text{cl.}}{\leq} C
		={}&
		\left[
			(\vec k+\vec l)!
			p_{\vec k+\vec l}
		\right]_{\vec k,\vec l\in \mathcal I}
		\\
		\text{and}\quad
		0\stackrel{\text{cl.}}{\leq} M
		={}&
		\left[
			\left\langle{:}
				\vec{\hat n}^{\vec k+\vec l}
			{:}\right\rangle
		\right]_{\vec k,\vec l\in \mathcal I},
	\end{aligned}
	\end{equation}
	where rows and columns of the matrices are determined by the multi-indices $\vec k=[k_1,\ldots,k_\mu]$ and $\vec l=[l_1,\ldots,l_\mu]$, respectively.
	In the above formulas, we also make use of the common multi-index operations $\vec n!=n_1!\cdots n_\mu!$ and $\vec{\hat n}^{\vec m}=\hat n_1^{m_1}\cdots\hat n_\mu^{m_\mu}$, as well as the joint photon-number counts $p_{\vec n}=p_{n_1,\ldots,n_\mu}$ for coincidences.

	Importantly, we can take any index set $\mathcal I\subset\mathcal I_1\times\cdots\times \mathcal I_\mu$ for the construction of $C$ and $M$.
	In addition, each component $\mathcal I_j$ ($j\in\{1,\ldots,\mu\}$) can be chosen---independently of all other modes---to consist of either integers $\mathbb N$ or half integers $\frac{1}{2}\mathbb N\setminus\mathbb N$ while one commonly only selects integer sets for all modes.
	With our approach, we thus have an exponential scaling of additional choices for matrices we can construct to assess nonclassicality.
	This results in $2^\mu$ distinct counting matrices $C$ and moment matrices $M$, depending on which of the two choices, integer or half-integer, we take for each mode.
	For the example of $\mu=2$ modes, we can have $\mathcal I_1\times\mathcal I_2$ being equal to $\mathbb N\times\mathbb N$, $\left(\frac{1}{2}\mathbb N\setminus\mathbb N\right)\times\mathbb N$, $\mathbb N\times\left(\frac{1}{2}\mathbb N\setminus\mathbb N\right)$, and $\left(\frac{1}{2}\mathbb N\setminus\mathbb N\right)\times\left(\frac{1}{2}\mathbb N\setminus\mathbb N\right)$.

\subsection{Multimode example}

	As an example, we consider generic $\mu$-mode even and odd coherent states,
	\begin{equation}
		\label{eq:MultimodeCat}
		|\vec\alpha_\pm\rangle
		=
		\frac{
			|\vec\alpha\rangle\pm|-\vec\alpha\rangle
		}{\sqrt{
			2\left(
				1\pm e^{-2\|\vec\alpha\|^2}
			\right)
		}},
	\end{equation}
	which are defined through the coherent amplitudes $\vec \alpha=[\alpha_j]_{j=1}^\mu\in\mathbb C^\mu$, with $\|\alpha\|^2=\sum_{j=1}^\mu|\alpha_j|^2$.
	Such states are interesting as they can approximate distinct forms of entanglement, such as GHZ states and W states for $\mu=3$ modes \cite{SV20}.

	The multimode, normally-ordered number moments for states in Eq. \eqref{eq:MultimodeCat} take the form
	\begin{equation}
		\langle{:}
			\vec{\hat n}^{\vec m}
		{:}\rangle
		=
		|\vec \alpha^{\vec m}|^2
		\frac{
			1
			\pm
			(-1)^{|\vec m|}
			e^{-2\|\vec\alpha\|^2}
		}{
			1\pm e^{-2\|\vec\alpha\|^2}
		},
	\end{equation}
	using the multi-index notation, where $\vec n=[n_j]_{j=1}^\mu\in\mathbb N^\mu$, $\vec\alpha^{\vec m}=\prod_{j=1}^\mu \alpha_j^{m_j}$, and $|\vec n|=\sum_{j=1}^\mu n_j$.
	For example, this results in the total photon number as
	\begin{equation}
		\label{eq:MultimodePhotonNumber}
		\langle\hat N\rangle
		=
		\sum_{j=1}^\mu\langle\hat n_j\rangle
		=
		\|\vec \alpha\|^2
		\left[
			\tanh(\|\vec \alpha\|^2)
		\right]^{\pm1}
	\end{equation}
	for the multimode even and odd coherent states, $|\vec\alpha_\pm\rangle$.
	Note that $\langle\vec\alpha_-|\hat N|\vec\alpha_-\rangle$ is lower-bounded by one because of $\lim_{\|\vec\alpha\|^2\to0}\|\vec\alpha\|^2/\tanh(\|\vec\alpha\|^2)=1$.

	As in the single-mode case, the simplest nonclassicality criteria, using a two-element index set $\mathcal I=\{\vec n,\vec m\}$, takes the form $0>
		\langle{:}
			\vec{\hat n}^{2\vec m}
		{:}\rangle\langle{:}
			\vec{\hat n}^{2\vec n}
		{:}\rangle-\langle{:}
			\vec{\hat n}^{\vec m+\vec n}
		{:}\rangle^2
	$.
	As a classical contstraint, this criterion may be recast into the form
	\begin{equation}
		\label{eq:ExampleMultimodeConstraint}
		1
		\stackrel{\text{cl.}}{\geq}
		\frac{
			\langle{:}
			\vec{\hat n}^{\vec m+\vec n}
			{:}\rangle^2
		}{
			\langle{:}
			\vec{\hat n}^{2\vec m}
			{:}\rangle
			\langle{:}
			\vec{\hat n}^{2\vec n}
			{:}\rangle
		}.
	\end{equation}
	The right-hand side can be computed for the multimode even and odd coherent states in Eq. \eqref{eq:MultimodeCat} as
	\begin{equation}
	\begin{aligned}
		{}&
		\frac{
			\langle{:}
			\vec{\hat n}^{\vec m+\vec n}
			{:}\rangle^2
		}{
			\langle{:}
			\vec{\hat n}^{2\vec m}
			{:}\rangle
			\langle{:}
			\vec{\hat n}^{2\vec n}
			{:}\rangle
		}
		\\
		={}&
		\frac{\left[
			1
			\pm
			(-1)^{|\vec m|+|\vec n|}
			e^{-2\|\vec\alpha\|^2}
		\right]^2}{
			\left[
				1
				\pm
				(-1)^{2|\vec m|}
				e^{-2\|\vec\alpha\|^2}
			\right]
			\left[
				1
				\pm
				(-1)^{2|\vec n|}
				e^{-2\|\vec\alpha\|^2}
			\right]
		},
	\end{aligned}
	\end{equation}
	In the context of half and full integer values of $\vec n$ and $\vec m$, we can then identify the following four cases:
		(i)
		We have integer values $|\vec n|,|\vec m|\in\mathbb N$ whose sum $|\vec n|+|\vec m|$ is even, e.g., $|\vec n|=0$ and $|\vec m|=2$.
		We find
		\begin{equation}
			\frac{
				\langle{:}
				\vec{\hat n}^{\vec m+\vec n}
				{:}\rangle^2
			}{
				\langle{:}
				\vec{\hat n}^{2\vec m}
				{:}\rangle
				\langle{:}
				\vec{\hat n}^{2\vec n}
				{:}\rangle
			}
			=
			1,
		\end{equation}
		  for the states considered here, which a trivial constant and never violates the inequality in Eq. \eqref{eq:ExampleMultimodeConstraint}.
		(ii)
		We have integer values $|\vec n|,|\vec m|\in\mathbb N$ whose sum $|\vec n|+|\vec m|$ is odd, e.g., $|\vec n|=0$ and $|\vec m|=1$.
		We find
		\begin{equation}
			\label{eq:MultimodeCase2}
			\frac{
				\langle{:}
				\vec{\hat n}^{\vec m+\vec n}
				{:}\rangle^2
			}{
				\langle{:}
				\vec{\hat n}^{2\vec m}
				{:}\rangle
				\langle{:}
				\vec{\hat n}^{2\vec n}
				{:}\rangle
			}
			=\left[
				\tanh(\|\vec\alpha\|^2)
			\right]^{\pm2},
		\end{equation}
		which is smaller than one for $|\vec\alpha_+\rangle$ and larger than one (i.e., nonclassical) for $|\vec\alpha_-\rangle$ because of $\tanh(\|\vec\alpha\|)^2<1$ and $1/\tanh(\|\vec\alpha\|)^2>1$, respectively.
		(iii)
		We have half-integer values $|\vec n|,|\vec m|\in\frac{1}{2}\mathbb N \setminus\mathbb N$ whose sum $|\vec n|+|\vec m|$ is even, e.g., $|\vec n|=1/2$ and $|\vec m|=3/2$.
		We find
		\begin{equation}
			\label{eq:MultimodeCase3}
			\frac{
				\langle{:}
				\vec{\hat n}^{\vec m+\vec n}
				{:}\rangle^2
			}{
				\langle{:}
				\vec{\hat n}^{2\vec m}
				{:}\rangle
				\langle{:}
				\vec{\hat n}^{2\vec n}
				{:}\rangle
			}
			=
			\left[
				\coth(\|\vec\alpha\|^2)
			\right]^{\pm 2},
		\end{equation}
		which certifies nonclassicality for $|\vec\alpha_+\rangle$ but not for $|\vec\alpha_-\rangle$, being the inverse of the previous case.
        (iv)
        We have half-integer values $|\vec n|,|\vec m|\in\frac{1}{2}\mathbb N \setminus\mathbb N$ whose sum $|\vec n|+|\vec m|$ is odd, e.g., $|\vec n|=1/2$ and $|\vec m|=5/2$.
		We find
		\begin{equation}
			\frac{
				\langle{:}
				\vec{\hat n}^{\vec m+\vec n}
				{:}\rangle^2
			}{
				\langle{:}
				\vec{\hat n}^{2\vec m}
				{:}\rangle
				\langle{:}
				\vec{\hat n}^{2\vec n}
				{:}\rangle
			}
			=
			1,
		\end{equation}
		which, similarly to the first case, presents a trivial scenario for the multimode states under study.

\begin{figure}
	\includegraphics[width=.8\columnwidth]{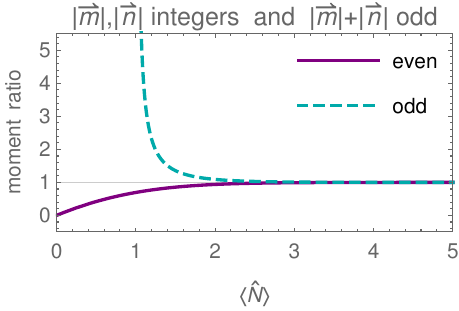}
	\\[2ex]
	\includegraphics[width=.8\columnwidth]{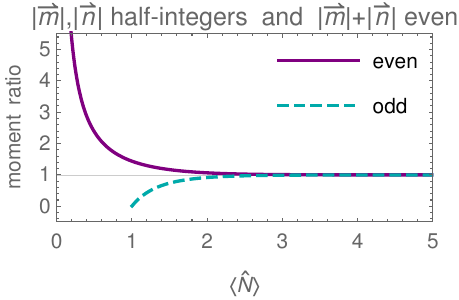}
	\caption{%
		Multimode nonclassicality witnessing in terms of moment ratios exceeding one, using the constraint in Eq. \eqref{eq:ExampleMultimodeConstraint} and states in Eq. \eqref{eq:MultimodeCat}, as a function of the total photon number $\langle\hat N\rangle=\langle\hat n_1\rangle+\cdots+\langle\hat n_\mu\rangle$.
		The top plot, pertaining to the case (ii) in Eq. \eqref{eq:MultimodeCase2}, certifies the nonclassicality of the multimode odd coherent state $|\vec\alpha_-\rangle$ through ratios above one for arbitrary amplitudes $\vec\alpha\in\mathbb C^\mu$ and arbitrary numbers of modes, $\mu$.
		The bottom plot, addressing the case (iii) in Eq. \eqref{eq:MultimodeCase3}, achieves the same result but for the multimode even coherent state $|\vec\alpha_+\rangle$.
		In the macroscopic limit, $\langle\hat N\rangle\to\infty$, we find that all curves converge to the classical boundary, i.e., the ratio one.
	}\label{fig:Multimode}
\end{figure}

	Figure \ref{fig:Multimode} depicts the verification of multimode nonclassicality via the ratio of moments in Eq. \eqref{eq:ExampleMultimodeConstraint}.
	In the top plot, we verify through values exceeding one that the aforementioned case (ii) renders it possible to detect nonclassical correlations in an odd superposition of tensor products of coherent states, $|\vec\alpha_-\rangle\propto\bigotimes_{j=1}^\mu|\alpha_j\rangle-\bigotimes_{j=1}^\mu|-\alpha_j\rangle$, but is insensitive to the even state.
	Conversely, the bottom plot confirms case (iii) as a successful choice for $|\vec\alpha_+\rangle\propto\bigotimes_{j=1}^\mu|\alpha_j\rangle+\bigotimes_{j=1}^\mu|-\alpha_j\rangle$ to certify its nonclassical correlations, while being ineffective for odd states.
	Without probing both the integer and half-integer criteria, half of the quantum correlations would remain unnoticed in this example.

	As a final remark, we note that the criteria based on multimode photon-number coincidence counts $p_{\vec m}=p_{m_1,\ldots,m_M}$ result in similar findings as obtained by the moments computed for the states in Eq. \eqref{eq:MultimodeCat}.
	Specifically, we have
	\begin{equation}
		\vec m!p_{\vec m}
		=
		\langle{:}
			\vec{\hat n}^{\vec m}
			e^{-|\vec{\hat n}|}
		{:}\rangle
		=
		|\vec \alpha^{\vec m}|^2
		\frac{
			1\pm(-1)^{|\vec m|}
		}{
			e^{\|\vec\alpha\|^2}\pm e^{-\|\vec\alpha\|^2}
		}.
	\end{equation}
	In terms of the simplest constraints, akin to Eq. \eqref{eq:ExampleMultimodeConstraint}, this allows us to write
	\begin{equation}
		1
		\stackrel{\text{cl.}}{\geq}
		\frac{\left[
			(\vec m+\vec n)!p_{\vec n+\vec m}
		\right]^2}{
			(2\vec m!)p_{2\vec m}
			(2\vec n!)p_{2\vec n}
		}
		=\frac{\left[
			1{\pm} (-1)^{|\vec m|{+}|\vec n|}
		\right]^2}{
			\left[
				1{\pm} (-1)^{2|\vec m|}
			\right]
			\left[
				1{\pm} (-1)^{2|\vec n|}
			\right]
		}.
	\end{equation}
	Analogous to the cases discussed for the multimode moments,
	for the odd state $|\vec\alpha_-\rangle$, $|\vec n|,|\vec m|\in\mathbb N$, and $|\vec n|+|\vec m|$ odd, the right-hand side formally reads $2^2/(0\cdot 0)=\infty$, which trivially exceeds the classical bound of one;
	the same result applies to the even state $|\vec\alpha_+\rangle$, $|\vec n|,|\vec m|\in\frac{1}{2}\mathbb N\setminus\mathbb N$, and $|\vec n|+|\vec m|$ even.
	(Note that the divergence $\infty$ does not show up for realistic efficiencies $0<\eta<1$.)

\section{Conclusion}
\label{Sec:Conclusion}

	To summarize, we constructed classes of single-mode and multimode nonclassicality witnesses for multiplexed click-counting detection schemes.
	Beyond the common integer-based criteria, the previously untapped potential of half-integer-based witnesses was made accessible.
	This resulted in an exponentially increased number of nonclassicality tests that can be directly harnessed in contemporary experiments.
	Realistic imperfections, including detector saturation, and arbitrary detector-response functions are included in our method.

	We showed that criteria using the integer and half-integer approaches are related to the even and odd photon-number parity of the state under study.
	We thus tested our methods using single-mode and multimode, even and odd coherent states to demonstrate this relation, even in the presence of significant losses.
	In addition to moment-based witnesses, families of criteria were derived that directly use the measured statistics of counts and coincidence counts to reveal local nonclassical effects and quantum features of coincidence counts.

	The methodology derived in this work also opens paths for further investigations.
	For example, click-based and phase-sensitive nonclassicality tests have been previously studied in theory and experiment \cite{SVA15,LSV15a,PSSB24}, which can be advanced through the counting matrices.
	The multimode, half-integer approach can provide new insights into nonclassical polarization effects, which were previously studied in the click-counting context, too \cite{SVA16,PSBS22}.
	Also, extensions of click-based quantum state tomography techniques \cite{LSV15,BTBSSV18,SPBTEWLNLGVASW20} appear appealing for future applications.
	Moreover, beyond common quantum correlations, conditional nonclassical correlations using multiplexing, such as reported in Ref. \cite{SBDBJDVW16}, can be advanced using the half-integer paradigm presented here.
    Lastly, our approach can be the basis for machine learning approaches, both by reliably labeling training data as well as by providing a witness to compare against \cite{GB20,GBWBZBA21,JKSBBSG26}.

\begin{acknowledgments}
	S.K., L.A., and J.S. acknowledge the financial support of the Deutsche Forschungsgemeinschaft (DFG) via the TRR 142/3 (Project No. 231447078, Subproject No. C10) and the QuantERA project QuCABOoSE.
	M.J. and M.G. are supported by funding from the German Research Foundation (DFG) under the project identifiers 398816777-SFB 1375 (NOA) and 550495627-FOR 5919 (MLCQS), from the Carl-Zeiss-Stiftung within the QPhoton Innovation Project MAGICQ, and from the Federal Ministry of Research, Technology and Space (BMFTR) under project BeRyQC.
\end{acknowledgments}

\end{document}